\documentclass[times,twocolumn,final,authoryear]{elsarticle}

\usepackage{amssymb}
\usepackage{aas_macros}
\usepackage{jasr}
\usepackage{framed,multirow}

\def\ion#1#2{#1\,{\sc #2}}

\newcommand{\ecss}{erg\,cm$^{-2}$\,s$^{-1}$\,sr$^{-1}$} 
\newcommand{\kms}{km\,s$^{-1}$}
\newcommand{\as}{${^\prime}{^\prime}$}
\newcommand{\hinode}{\textit{Hinode}}

\usepackage{url}
\usepackage{xcolor}
\definecolor{newcolor}{rgb}{.8,.349,.1}

\usepackage[citebordercolor=white]{hyperref}

\journal{Advances in Space Research}

\begin{document}

\verso{Peter Young \textit{etal}}

\begin{frontmatter}



\title{A Spectroscopic Measurement of High Velocity Spray Plasma from an M-class Flare and Coronal Mass Ejection}%


\author[1,2]{Peter R. \snm{Young}}

\address[1]{NASA Goddard Space Flight Center, Code 671, Heliophysics Division, Greenbelt, MD 20771, USA}
\address[2]{Department of Mathematics, Physics and Electrical Engineering, Northumbria University,
Newcastle upon Tyne, UK}

\begin{abstract}
Coronal mass ejection spray plasma associated with the M1.5-class flare of 16 February 2011 is found to exhibit a Doppler blue-shift of 850~\kms\ -- the largest value yet reported from ultraviolet (UV) or extreme ultraviolet (EUV) spectroscopy of the solar disk and inner corona. The observation is unusual in that the emission line (\ion{Fe}{xii} 193.51~\AA) is not observed directly, but the Doppler shift is so large that the blue-shifted component appears in a wavelength window at 192.82~\AA, intended to observe lines of \ion{O}{v}, \ion{Fe}{xi} and \ion{Ca}{xvii}. The \ion{Fe}{xii} 195.12~\AA\ emission line is used as a proxy for the rest component of 193.51~\AA. The observation highlights the risks of using narrow wavelength windows for spectrometer observations when observing highly-dynamic solar phenomena. The consequences of large Doppler shifts for ultraviolet solar spectrometers, including the upcoming Multi-slit Solar Explorer (MUSE) mission, are discussed.
\end{abstract}

\begin{keyword}
\KWD Solar atmosphere\sep Solar flares\sep Solar corona\sep Solar extreme ultraviolet emission\sep Solar coronal mass ejections\sep Spectroscopy
\end{keyword}

\end{frontmatter}



\section{Introduction}

Active region AR 11158 produced the first X-class flare of Solar Cycle 24 on 15 February 2011 \citep{2011ApJ...738..167S}. This was the most intense of  many C and M-class flares between 13 and 18 February. The M1.6 eruptive flare that peaked at 14:25~UT on February 16  was a rare example of an extreme ultraviolet (EUV) wave captured with a sit-and-stare spectroscopic study of the \textit{EUV Imaging Spectrometer} \citep[EIS:][]{2007SoPh..243...19C}. The observation was  run as part of \textit{Hinode} Operation Plan No.~180, organized by P.~G\"om\"ory and A.~Veronig. Four previous articles have presented the EIS results from the event (detailed below), but none of them noted a plasma component that had a very large Doppler velocity -- perhaps the largest ever recorded in the inner corona -- associated with ejected coronal plasma.

In this article I refer to the ejecta as \textit{CME spray} (CME: coronal mass ejection), preferring it to \textit{flare spray} -- a term that was introduced by \citep{1957ApJ...125..811W} prior to the discovery of CMEs. The latter authors distinguished spray plasmas from plasmas associated with erupting prominences seen at the limb, highlighting their clumpiness and large speeds. Modern interpretations of CMEs identify the erupting prominence as one part of a flux rope that is ejected from the Sun \citep{2006JGRA..11112103G}. The flux rope may contain coronal plasma \citep{2011ApJ...732L..25C}, and the CME spray could be part of this structure, or it could be ambient active region plasma blasted out by the ejection of the flux rope.





The EUV wave produced by the 16 February event was identified from EUV running-difference images obtained with the \textit{Atmospheric Imaging Assembly} \citep[AIA:][]{2012SoPh..275...17L} on board the \textit{Solar Dynamics Observatory}.  
\citet{2011ApJ...737L...4H} and \citet{2011ApJ...743L..10V} found that the EUV wave front was co-spatial with a region of red-shifts identified in coronal emission lines and moving along the EIS slit at speeds around 500~\kms. Immediately following this was a region of weak blue-shifts and the interpretation was of the EUV wave pushing down on the low-lying corona as it moved outward, generating red-shifts, followed by a weak rebound of the corona yielding the blue-shifts. EIS data for the flare associated with the EUV wave were studied in detail by \citet{2016A&A...588A...6G}, while \citet{2013SoPh..288..567L} used the wave kinematics measured from AIA and density measurements from EIS to estimate the magnetic field strength of the disturbed quiet corona. 

\citet{2011ApJ...737L...4H} highlighted that strong blue-shifted secondary components to \ion{Fe}{xii} 195.12~\AA\ are seen early in the flare, and noted that ``the secondary component is so fast that it falls out of the EIS spectral window in some cases," implying speeds greater than 400~\kms. \citet{2016A&A...588A...6G} also noted a  blue-shifted component  of more than 450~\kms\ for the \ion{Fe}{xiii} 202.04~\AA\ line, but cautioned that the feature may actually be due to another emission line. In both cases these high-speed blue-shifted components are close to the flare site.

In this article I demonstrate that the blue-shifted feature of \ion{Fe}{xii} actually reaches speeds of 850~\kms\ and moves along the EIS slit, tracking behind the EUV wavefront. It corresponds to coronal spray plasma released by the CME and seen in the AIA 193~\AA\ channel.

Software and data files used for generating the figures and table for this article are available in the GitHub repository \url{https://github.com/pryoung/papers/tree/main/2022_eis_spray}, which also contains two mp4-format movies. In the following text the repository will be referred to as \textsf{2022-eis-spray}.






\begin{figure*}[t]
    \centering
    \includegraphics[width=\textwidth]{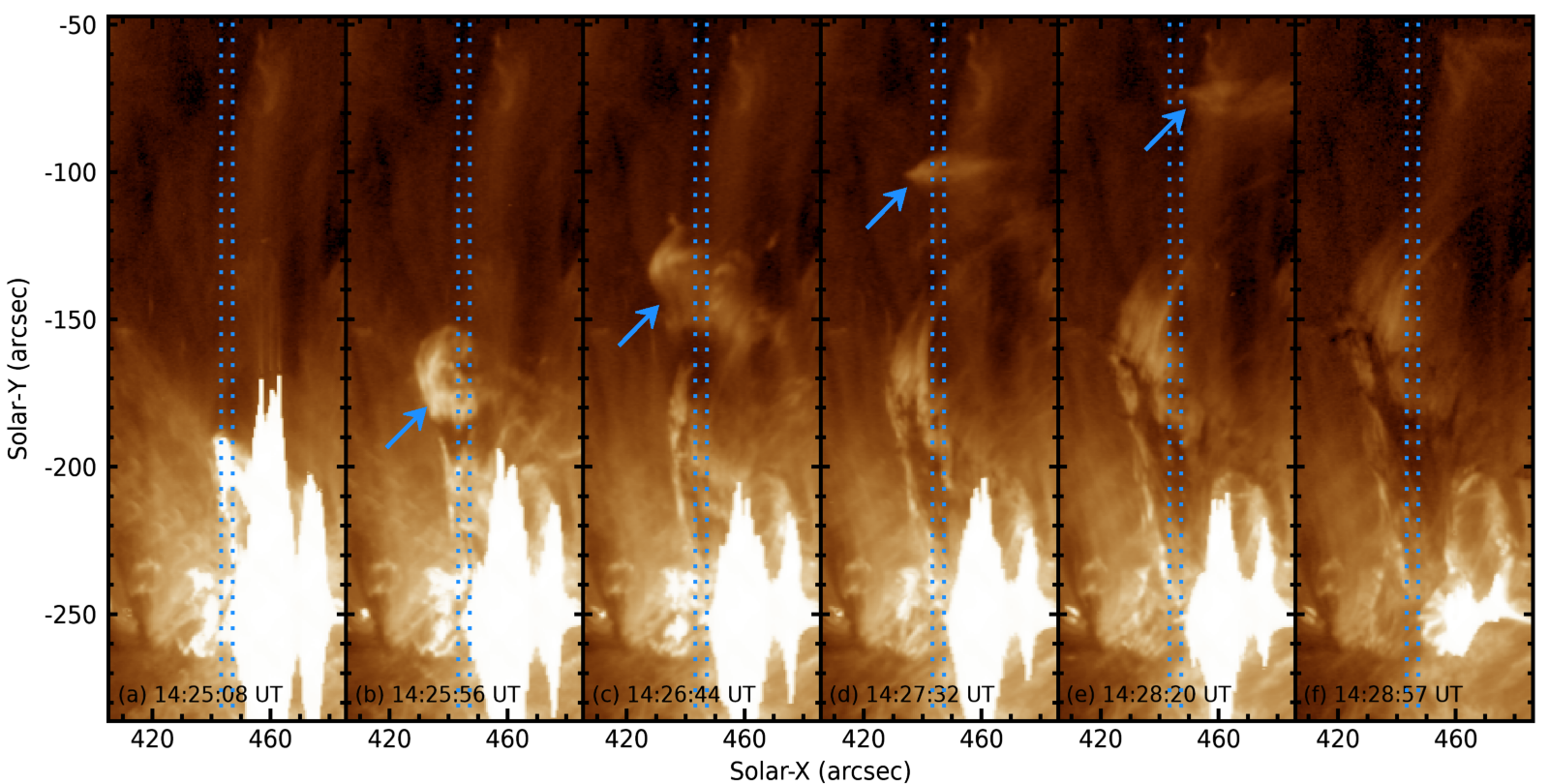}
    \caption{Six AIA 193~\AA\ image frames showing the spray plasma ejected from the flaring active region. \textit{Vertical dotted blue lines} show the location of the EIS slit, and \textit{blue arrows} identify the plasma component that gives rise to the high Doppler velocities measured by EIS. A logarithmic intensity scaling is used for each frame.}
    \label{fig.aia}
\end{figure*}

\section{Observations}

The  study \textsf{eis:eitwave\_obssns} (designed by P.~G\"om\"ory) was used for the EIS observation, and it used the 2\as\ slit in the sit-and-stare mode. The exposure time was 45~s, and the cadence averaged 46.9~s. Eleven wavelength windows were downloaded, each of width 24 pixels. Of interest for the present work are the windows centered at 192.82 and 195.12~\AA. The former contains a complex of lines including six lines of \ion{O}{v}, \ion{Fe}{xi} 192.81~\AA\ (usually the strongest line) and \ion{Ca}{xvii} 192.85~\AA, which can become strong during flares. 
The 195.12~\AA\ window is centered on an \ion{Fe}{xii} line that is usually the strongest line observed by EIS in terms of number of counts. The 24-pixel widths of the windows mean that the maximum Doppler shift that can be measured for a line centered in a window is 410~\kms.

\ion{Fe}{xii} gives rise to another strong line at 193.51~\AA\ that is two-thirds the strength of the line at 195.12~\AA. It is rarely observed as observers prefer to use the stronger line, and there is no diagnostic benefit to observing both lines. If 193.51~\AA\ shows a large enough blue-shift, however, then it can make an appearance in the 192.82~\AA\ window. This is what occurred for the 16 February flare.

\section{Results}

\begin{figure*}[t]
    \centering
    \includegraphics[width=\textwidth]{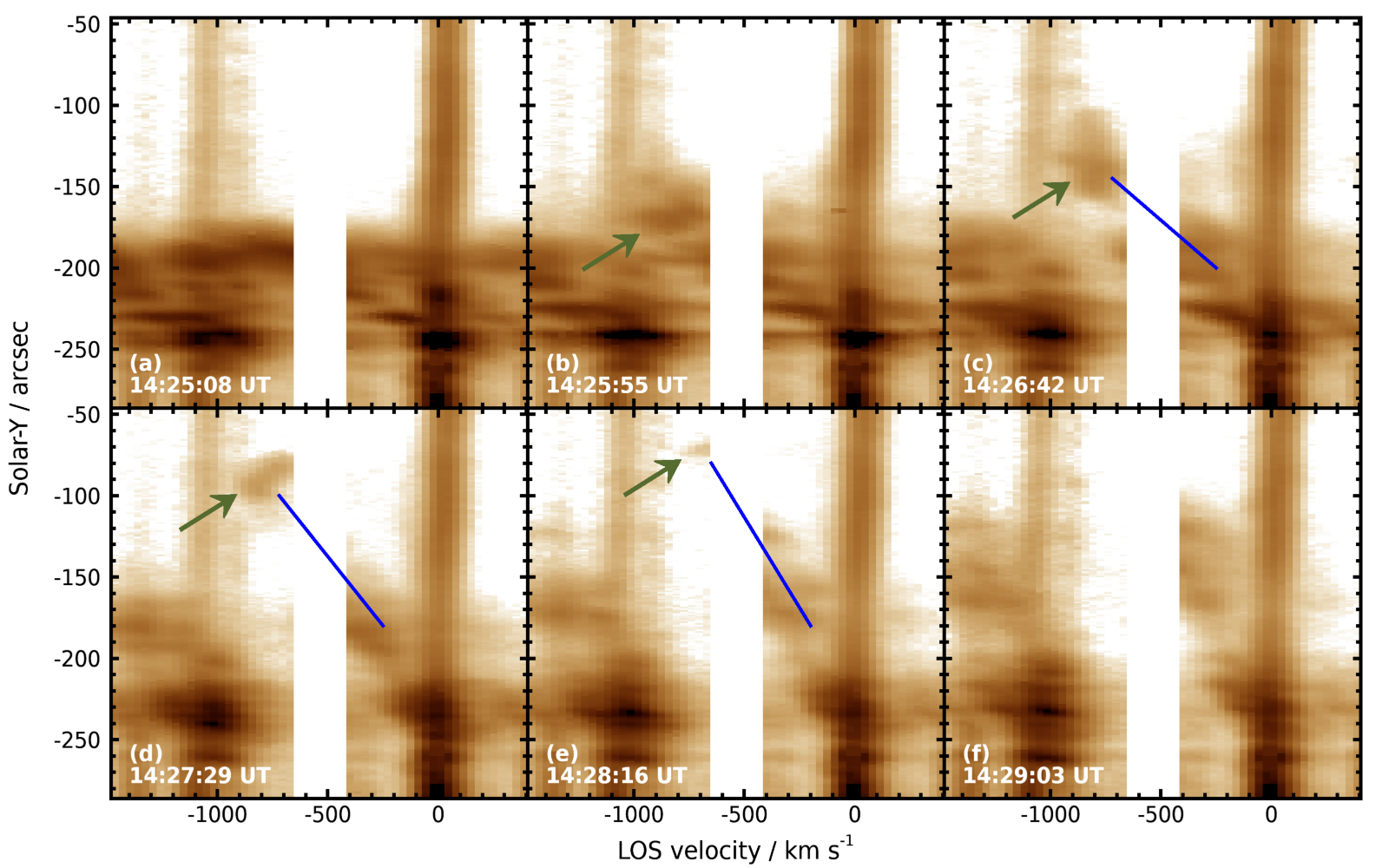}
    \caption{Six consecutive exposure pairs that show the evolution of the high-velocity spray plasma (indicated by \textit{green arrows}). Each panel shows the 192.82~\AA\ (left) and 195.12~\AA\ (right) wavelength windows. They are placed in velocity space (x-axis) such that the 195.12~\AA\ window is positioned at 193.51~\AA\ and scaled by a factor 0.675. \textit{Blue diagonal lines} highlight the connection between the high-velocity and lower-velocity spray plasmas seen in the two windows. Images are shown with inverse-logarithmic scaling. }
    \label{fig.fe12-spec}
\end{figure*}

Six AIA 193~\AA\ image frames are shown in Figure~\ref{fig.aia} that show the CME spray ejected northwards from the flare site (identified by the saturated emission at the bottom of the images). The frames were selected to correspond to the mid-time of the six EIS exposures shown in Figure~\ref{fig.fe12-spec}. \textit{Vertical blue dotted lines} indicate the location of the EIS slit, as determined by \citet{2016A&A...588A...6G}. \textit{Blue arrows} on frames (b) through (e) identify the plasma component that gives rise to the high-velocity Doppler component seen by EIS. An mp4 movie available in the \textsf{2022-eis-spray} repository shows the full-cadence AIA 193~\AA\ image sequence, with the approximate location of the EIS slit over-plotted.

\begin{table*}
\centering
\caption{Gaussian fit parameters for \ion{Fe}{xii} Doppler components, and derived blue-shifts.\label{tbl.eis}}
\begin{tabular}{|c|c|c|c|c|c|c|}
\hline
Exposure    & y-pixels & $I_\mathrm{spray}^a$ & $I_\mathrm{rest}^a$ & $\lambda_\mathrm{spray}$ (\AA)& $\lambda_\mathrm{rest}$ (\AA) & $v$ (\kms)\\
\hline
14:26:42 & 158:165 & 145 & 145  & 192.981 & 193.518 & $-831$        \\
\hline
14:27:29 & 197:204 & 58 & 148  & 192.982 & 193.528 & $-847$        \\
\hline
\multicolumn{7}{l}{$^a$ Units: \ecss.}
\end{tabular}
\end{table*}

Figure~\ref{fig.fe12-spec} shows the development of the CME spray plasma as seen with EIS. The panels (a) to (f) correspond to six consecutive exposures from the EIS data file beginning at 14:09:55~UT. The time shown for each panel is the mid-point time of the 45~s exposures. A single EIS exposure has wavelength in the x-direction and solar-y position in the y-direction. Each panel of Figure~\ref{fig.fe12-spec} combines the 192.82~\AA\ (left) and 195.12~\AA\ (right) wavelength windows. Since the \ion{Fe}{xii} 195.12~\AA\ emission line should behave exactly as \ion{Fe}{xii} 193.51~\AA, I have placed the two windows in such a way that the 195.12~\AA\ line is at the position of the 193.51~\AA\ line. The 195.12~\AA\ window intensity has been multipled by a factor 0.675 to account for the expected ratio of the two \ion{Fe}{xii} lines. The wavelength axis has been converted to line-of-sight (LOS) velocity using 193.51~\AA\ as the rest wavelength. The 192.82~\AA\ window then occupies the velocity region $-1480$ to $-690$~\kms. The \ion{Fe}{xi} 192.81~\AA\ line, which can be seen extending the full height of the window, is located at around $-1050$~\kms. An mp4 movie showing the full set of 40 exposure-pairs from the EIS data file is available in the \textsf{2022-eis-spray} repository.

Panel (a) shows complex structure between $y=-250$\as\ and $y=-180$\as. By comparing the two windows, common structures can be seen, implying similar velocity structures in the \ion{Fe}{xi} and \ion{Fe}{xii} lines. At around $y=-200$\as\ to $y=-180$\as\ there appears to be both red and blue-shifted plasma components, and a rest component to the lines, giving a band of emission across the exposures. It is possible that some of this emission in the 192.82~\AA\ window corresponds to a high-velocity component of \ion{Fe}{xii} 193.51~\AA\ around $-800$ to $-700$~\kms\ but comparisons with the cool \ion{Si}{vii} 275.36~\AA\ and \ion{Mg}{vii} 278.39~\AA\ lines also seen by EIS suggest this is more likely to be cool \ion{O}{v} 192.90~\AA\ emission combined with \ion{Fe}{xi}.

Panel (b) is the first to show clear evidence of a high-velocity \ion{Fe}{xii} component, which is indicated with a green arrow. None of the other wavelength windows show emission at this location, i.e., on the long-wavelength side of the central emission line. The position and time matches the spray plasma seen with AIA (Figure~\ref{fig.aia}b). This component is more clearly seen in panels (c) and (d) -- again marked with green arrows -- as it moves along the slit. The final appearance is in panel (e). The velocity has decreased resulting in only a small part of the emission appearing in the 192.82~\AA\ window.

For the exposures at 14:26:41~UT and 14:27:29~UT, where the high-velocity Doppler component is most clearly seen, Gaussian fits were performed in order to determine the velocity of the spray plasma. A range of y-pixels were averaged for each exposure using the IDL routine \textsf{eis\_mask\_spectrum} \citep{2022zndo...6339584Y} to yield a 1D spectrum. Gaussian fits were performed with the routine \textsf{spec\_gauss\_eis} \citep{2022zndo...6339584Y}  to the features in the 192.82 and 195.12~\AA\ windows, yielding intensities and centroids for the spray component (measured in the 193.82~\AA\ window) and the ``rest'' component (measured in the 195.12~\AA\ window). The intensities of the rest components were multiplied by 0.675 to yield the estimated 193.51~\AA\ intensities, and the centroids of the rest components were reduced by 1.610~\AA. The latter two values were both obtained from the CHIANTI database \citep{2016JPhB...49g4009Y,2021ApJ...909...38D}. The intensity ratio was obtained assuming a temperature of $10^{6.2}$~K and electron number density of $10^9$~cm$^{-3}$, although the ratio shows only a small variation with respect to these parameters. The spray Doppler shifts (relative to the measured rest components) for the two exposures are $-831$ and $-847$~\kms.

\begin{figure}[t]
    \centering
    \includegraphics[width=3in]{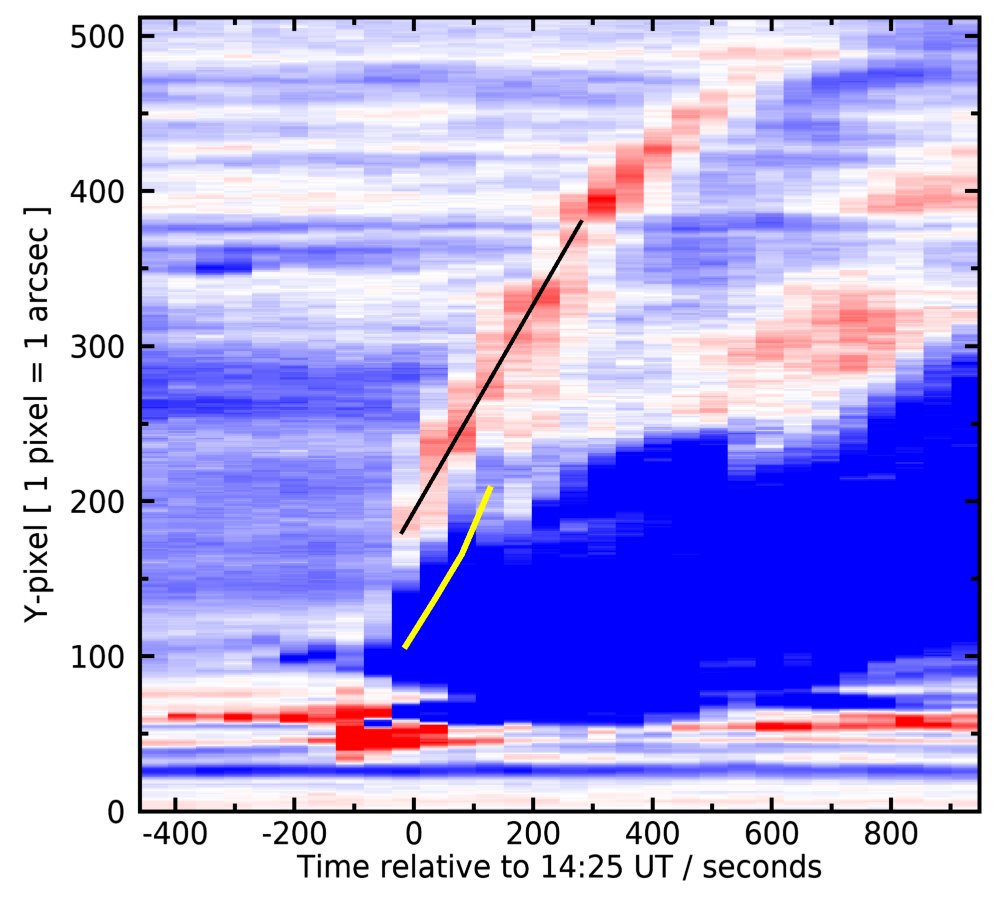}
    \caption{Doppler shifts of the \ion{Fe}{xii} 195.12~\AA\ line as a function of time, given relative to the time of peak X-ray flux (14:25~UT). The image is scaled between $-15$ and $+15$~kms. The \textit{black diagonal line} marks the weak red-shifts associated with the EUV wave, and the \textit{yellow diagonal line} marks the centroids of the high-velocity spray component. }
    \label{fig.fe12-vel}
\end{figure}

Following behind the high-velocity component there is further spray plasma emission that has a smaller Doppler shift, and the location in relation to the high-velocity component is indicated by the diagonal blue lines in panels (c) to (e). If the spectral region in the gap between the two wavelength windows had been available we would likely have seen a continuum of emission between the two velocity components. Note that the lower-velocity plasma seen in the 195.12~\AA\ emission continues to move northwards along the slit in panel (f).


Is there an alternative explanation for the emission seen in the 192.82~\AA\ window? The feature could be due to \ion{O}{v} 192.90~\AA, which would require a  red-shift of around 100~\kms. Other cool lines observed by EIS are \ion{He}{ii} 256.32~\AA\ and \ion{Si}{vii} 275.36~\AA, but neither of these show evidence of a similar red-shift. The feature could be a blue-shifted component of a line at a shorter wavelength than 193.51~\AA. However, the spectral atlas of \citet{2008ApJS..176..511B} shows that there are no strong lines between 193.51~\AA\ and the 192.82~\AA\ complex that could account for the relatively strong signal of the spray plasma.

Figure~\ref{fig.fe12-vel} shows a Dopplergram formed through a single-Gaussian fit to the \ion{Fe}{xii} 195.12~\AA\ emission line. The fit was performed with the IDL routine \textsf{eis\_auto\_fit} \citep{2022zndo...6339584Y} and was constrained to avoid fitting the extended wing on the short-wavelength side as this shows significant structure, particularly in the continuous blue region beginning at the flare peak. The image is comparable to the Dopplergrams shown in Figures~2 and 3 of \citet{2011ApJ...743L..10V} and \citet{2011ApJ...737L...4H}, respectively, although these authors used the slightly hotter \ion{Fe}{xiii} 202.04~\AA\ line. The narrow, red feature running bottom-left to to-right is the weak red-shifted component that \citet{2011ApJ...743L..10V} and \citet{2011ApJ...737L...4H} identified with the EUV wavefront. The gradient can be seen to decrease with time, marking a deceleration of the EUV wave. A straight line is marked on the early part of the red-shifted component and this corresponds to a speed of 480~\kms. Centroids in the y-direction of the spray plasma were estimated from the images shown in panels (b) to (e) of Figure~\ref{fig.fe12-spec}, and these are indicated by the yellow line on Figure~\ref{fig.fe12-vel}. The gradient is 515~\kms, but given the uncertainties due to the irregular shape of the spray emission I consider this to be consistent with the propagation speed of the red-shifted component. Assuming the latter for the plane-of-sky speed of the spray plasma (since the value is more accurate), the maximum Doppler shift speed of Table~\ref{tbl.eis} then implies an actual speed of the spray plasma of 975~\kms.

\begin{figure}[t]
    \centering
    \includegraphics[width=\columnwidth]{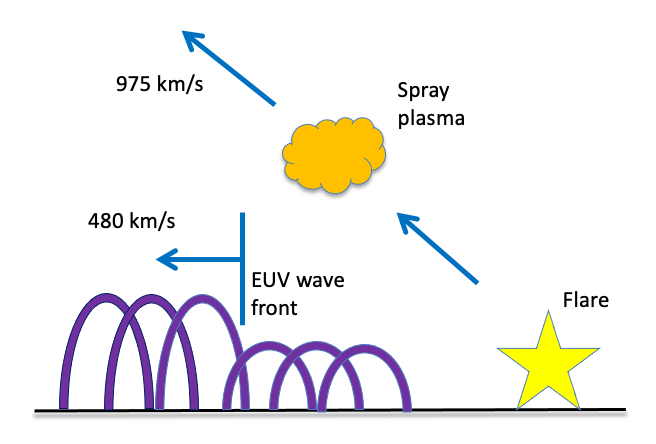}
    \caption{A cartoon illustrating the dynamics associated with the EUV wave and spray ejecta.}
    \label{fig.cartoon}
\end{figure}

Figure~\ref{fig.cartoon} shows my interpretation of the AIA and EIS data from the 16 February event. The star shows the location of the flare, and the loops denote quiet Sun coronal loops northward of the flare site. The EUV wave presses down on the quiet Sun as it moves outwards, leading to a propagating front of red-shifts \citep{2011ApJ...743L..10V,2011ApJ...737L...4H} that moves at around 480~\kms. The fastest spray plasma is ejected from the flare and tracks behind the EUV wave front by around 50~Mm, at a  similar plane-of-sky speed. The maximum line-of-sight speed is 850~\kms. Combining these two velocities implies the speed of the spray plasma is 975~\kms.

\section{Previous Measurements of Large Doppler Shifts}

Spectroscopic measurements of sprays have been made previously with CDS and EIS. As with the example presented here, the spray is revealed through a distinct blue-shifted component to the line profile that is seen in addition to a component close to the rest wavelength of the line (which likely is due to background plasma emission). \citet{2012ApJ...748..106T} referred to this as \textit{line-splitting}.  

\citet{2001ApJ...560L..91F} and  \cite{2002SoPh..206..359P} both studied spray ejecta from an X-class flare observed by CDS on 10 April 2001. They  reported blue-shifts of around 400~\kms\ from the \ion{O}{v} 629.7~\AA\ line formed at 0.2~MK. \citet{2003ApJ...587..429H} found blue-shifts in the same \ion{O}{v} line of up to 350~\kms\ for spray ejecta observed on 13 June 1998. There was evidence that the spray was also seen in the coronal \ion{Mg}{x} 624.9~\AA\ line, too.

The first EIS measurement of large Doppler shifts associated with flare ejecta was from \citet{2008ApJ...685..622A} who reported blue-shifted components at around 250~\kms\ for a wide range of EIS lines. This was associated with the X-class flare of 13 December 2007. A flare on 14 February 2011 (two days prior to the event discussed here) was studied by \citet{2012ApJ...748..106T} who found blue-shifted components up to 200~\kms, again in a range of EIS lines.

A joint observation of the X1-class flare on 29 March 2014 with EIS and the \textit{Interface Region Imaging Spectrograph} \citep[IRIS:][]{2014SoPh..289.2733D} was reported by \citet{2015ApJ...806....9K}. The filament eruption was captured with both EIS and IRIS, and the largest blue-shift found from IRIS was around 550~\kms\ and seen in lines of \ion{Mg}{ii}, \ion{C}{ii} and \ion{Si}{iv}. EIS observed the eruption at a slightly earlier phase when the blue-shift was smaller at around 220~\kms\ and the emission was seen in the cool \ion{He}{ii} ion as well as \ion{Fe}{xii}.

Coronal mass ejections observed well above the limb with the \textit{Ultraviolet Coronagraph Spectrometer} \citep[UVCS:][]{1995SoPh..162..313K} could yield very large Doppler shifts, as reported by \citet{2006ApJ...652..774C}. As the CME core (often associated with an erupted prominence) passed through the spectrometer slit, compact knots or threads would be seen. These features are typically Doppler-shifted, indicating that the plasma is moving towards or away from the observer as the CME moves outwards. Five examples of Doppler shifts greater than 800~\kms\ were found, with the largest a blue-shift of 1200~\kms. The measurements were made at heliocentric radii between 1.7 and 3.8~$R_\odot$.


Explosive chromospheric evaporation in the early stages of flares is another  solar phenomenon that can produce large Doppler shifts. A high energy input (usually in the form of electron beams) from the corona heats the chromosphere and the heated plasma rises into the corona. When observed by UV or X-ray spectrometers, a blue-shifted component to emission lines is seen, typically for temperatures of 1~MK and higher. The magnitude of the blue-shifts increases with temperature, and speeds up to around 400~\kms\ have been reported from EIS observations of \ion{Fe}{xxiii} and \ion{Fe}{xxiv} lines formed at 15--20~MK \citep{2010ApJ...719..213W,2013ApJ...766..127Y}.

Observations of flares above the limb have revealed another type of high-velocity feature that has only been seen in ions formed around 10~MK or hotter. In contrast to sprays and evaporation flows, the line profiles generally show a very broad wing emission that can be on the short or long-wavelength side of the profile, or both sides simultaneously. The first such measurement was presented by \citet{2001ApJ...549L.249I} using the \ion{Fe}{xx} 721.7~\AA\ line, which showed a blue-wing extending to 650~\kms. A similar profile was reported by \citet{2003SoPh..217..267I} for an X-flare observed on 21 April 2002, but for \ion{Fe}{xxi} 1354.1~\AA. The wing in this case extended for almost 5~\AA, corresponding to 1000~\kms. \citet{2007ApJ...661L.207W} studied the \ion{Fe}{xix} 1118.1~\AA\ line for an M3-class flare observed on 16 April 2002, which showed the blue wing extending to 600~\kms.

\citet{2013ApJ...776L..11I} presented similar line profiles from EIS for the \ion{Fe}{xxiv} 192.1~\AA\ emission line, observed from an X2-class flare on 27 January 2012. The wavelength window restricted velocities to $\pm$~400~\kms\ and, from the plots presented in the article, it is clear the blue-wing extends beyond this limit.

In summary, to the best of my knowledge, the Doppler-shifted component of 847~\kms\ found here is the largest reported in the literature from UV and EUV instruments for a distinct plasma component in the inner corona within 1.5~$R_\odot$.

\section{Summary and discussion}

The M2-class flare that peaked at 14:25~UT on 16 February 2011 was associated with a CME and an EUV wave. In the present article the fastest-moving ejecta observed with the \hinode/EIS instrument was found to have a Doppler shift of 850~\kms\ that I believe is the largest ever recorded by an ultraviolet spectrometer within 1.5~$R_\odot$. The plane-of-sky speed was 480~\kms, consistent with the speed of the EUV wave front. Combining the Doppler and plane-of-sky speeds gives a speed of 975~\kms.

The means of detecting the large Doppler shift is unusual in that the high-speed Doppler component was detected in an emission line that was not intended to be observed. The \ion{Fe}{xii} 193.51~\AA\ line was blue-shifted to such an extent that it appeared in a wavelength window centered at 192.82~\AA.

This observation highlights a limitation imposed on a UV spectrometer by the telemetry rate assigned to the mission. In order to achieve a good cadence and spatial coverage, only portions of the detector are downloaded that are typically centered on a selection of diagnostically-useful lines. In the case of EIS, wavelength windows would need to be around 60 pixels wide (compared to 24 pixels for the 16 February observation) to enable Doppler velocities of $\pm 1000$~\kms\ to be measured. Compared to the 24 pixel windows used for the 16 February dataset, this would be a 150\%\ increase in data volume. 

The presence of very large Doppler shifts may be a concern for the Multi-slit Solar Explorer \citep[MUSE:][]{2020ApJ...888....3D}, which is a novel spectrometer design recently selected by NASA for flight in 2026. Rather than a single slit, MUSE will have 37 parallel slits, which means that spectra from each slit will overlap on the detector. MUSE will have three channels for the \ion{Fe}{xix} 108.36~\AA, \ion{Fe}{ix} 171.07~\AA, and \ion{Fe}{xv} 284.16~\AA\ lines. The inter-slit spacing corresponds to 0.39~\AA\ for the two shorter wavelengths and 0.78~\AA\ for the longer. These translate to Doppler shifts of 1080, 680 and 820~\kms, respectively, for the three channels. Thus a large Doppler shift such as the one found in the present article may be confused with emission from a neighboring slit position. A benefit of MUSE, however, is that there is no equivalent of the windows used for EIS, and thus the complete Doppler range of $\pm 1000$~\kms\ (for example) will always be observed and so there is no risk of MUSE becoming ``blind" above a certain velocity limit.

\section*{Acknowledgments}
The author acknowledges support from  the  GSFC Internal Scientist Funding Model competitive work package program and the Heliophysics Guest Investigator program. J.T.~Karpen is thanked for valuable comments. \hinode\ is a Japanese mission developed
and launched by ISAS/JAXA, with NAOJ as domestic
partner and NASA and STFC (UK) as international partners.
It is operated by these agencies in co-operation with ESA and
NSC (Norway). 

 \bibliographystyle{jasr-model5-names} 
 \bibliography{ms}

\begin{thebibliography}{30}
\expandafter\ifx\csname natexlab\endcsname\relax\def\natexlab#1{#1}\fi
\ifx\xfnm\relax \def\xfnm[#1]{\unskip,\space#1}\fi

\bibitem[{{Asai} et~al.(2008){Asai}, {Hara}, {Watanabe}, {Imada}, {Sakao},
  {Narukage}, {Culhane} \& {Doschek}}]{2008ApJ...685..622A}
\bibinfo{author}{{Asai}, A.}, \bibinfo{author}{{Hara}, H.},
  \bibinfo{author}{{Watanabe}, T.} et~al. (\bibinfo{year}{2008}).
\newblock \bibinfo{title}{{Strongly Blueshifted Phenomena Observed with Hinode
  EIS in the 2006 December 13 Solar Flare}}.
\newblock {\it \bibinfo{journal}{\apj}\/},  {\it
  \bibinfo{volume}{685}\/}\bibinfo{issue}{(1)}, \bibinfo{pages}{622--628}.
  \DOIprefix\doi{10.1086/590419}. \href{http://arxiv.org/abs/0805.4468}{\tt
  arXiv:0805.4468}.

\bibitem[{{Brown} et~al.(2008){Brown}, {Feldman}, {Seely}, {Korendyke} \&
  {Hara}}]{2008ApJS..176..511B}
\bibinfo{author}{{Brown}, C.~M.}, \bibinfo{author}{{Feldman}, U.},
  \bibinfo{author}{{Seely}, J.~F.} et~al. (\bibinfo{year}{2008}).
\newblock \bibinfo{title}{{Wavelengths and Intensities of Spectral Lines in the
  171-211 and 245-291 {\r{A}} Ranges from Five Solar Regions Recorded by the
  Extreme-Ultraviolet Imaging Spectrometer (EIS) on Hinode}}.
\newblock {\it \bibinfo{journal}{\apjs}\/},  {\it
  \bibinfo{volume}{176}\/}\bibinfo{issue}{(2)}, \bibinfo{pages}{511--535}.
  \DOIprefix\doi{10.1086/529378}.

\bibitem[{{Cheng} et~al.(2011){Cheng}, {Zhang}, {Liu} \&
  {Ding}}]{2011ApJ...732L..25C}
\bibinfo{author}{{Cheng}, X.}, \bibinfo{author}{{Zhang}, J.},
  \bibinfo{author}{{Liu}, Y.} et~al. (\bibinfo{year}{2011}).
\newblock \bibinfo{title}{{Observing Flux Rope Formation During the Impulsive
  Phase of a Solar Eruption}}.
\newblock {\it \bibinfo{journal}{\apjl}\/},  {\it
  \bibinfo{volume}{732}\/}\bibinfo{issue}{(2)}, \bibinfo{pages}{L25}.
  \DOIprefix\doi{10.1088/2041-8205/732/2/L25}.
  \href{http://arxiv.org/abs/1103.5084}{\tt arXiv:1103.5084}.

\bibitem[{{Ciaravella} et~al.(2006){Ciaravella}, {Raymond} \&
  {Kahler}}]{2006ApJ...652..774C}
\bibinfo{author}{{Ciaravella}, A.}, \bibinfo{author}{{Raymond}, J.~C.},  \&
  \bibinfo{author}{{Kahler}, S.~W.} (\bibinfo{year}{2006}).
\newblock \bibinfo{title}{{Ultraviolet Properties of Halo Coronal Mass
  Ejections: Doppler Shifts, Angles, Shocks, and Bulk Morphology}}.
\newblock {\it \bibinfo{journal}{\apj}\/},  {\it
  \bibinfo{volume}{652}\/}\bibinfo{issue}{(1)}, \bibinfo{pages}{774--792}.
  \DOIprefix\doi{10.1086/507171}.

\bibitem[{{Culhane} et~al.(2007){Culhane}, {Harra}, {James}, {Al-Janabi},
  {Bradley}, {Chaudry}, {Rees}, {Tandy}, {Thomas}, {Whillock}, {Winter},
  {Doschek}, {Korendyke}, {Brown}, {Myers}, {Mariska}, {Seely}, {Lang}, {Kent},
  {Shaughnessy}, {Young}, {Simnett}, {Castelli}, {Mahmoud}, {Mapson-Menard},
  {Probyn}, {Thomas}, {Davila}, {Dere}, {Windt}, {Shea}, {Hagood}, {Moye},
  {Hara}, {Watanabe}, {Matsuzaki}, {Kosugi}, {Hansteen} \&
  {Wikstol}}]{2007SoPh..243...19C}
\bibinfo{author}{{Culhane}, J.~L.}, \bibinfo{author}{{Harra}, L.~K.},
  \bibinfo{author}{{James}, A.~M.} et~al. (\bibinfo{year}{2007}).
\newblock \bibinfo{title}{{The EUV Imaging Spectrometer for Hinode}}.
\newblock {\it \bibinfo{journal}{\solphys}\/},  {\it
  \bibinfo{volume}{243}\/}\bibinfo{issue}{(1)}, \bibinfo{pages}{19--61}.
  \DOIprefix\doi{10.1007/s01007-007-0293-1}.

\bibitem[{{De Pontieu} et~al.(2020){De Pontieu}, {Mart{\'\i}nez-Sykora},
  {Testa}, {Winebarger}, {Daw}, {Hansteen}, {Cheung} \&
  {Antolin}}]{2020ApJ...888....3D}
\bibinfo{author}{{De Pontieu}, B.}, \bibinfo{author}{{Mart{\'\i}nez-Sykora},
  J.}, \bibinfo{author}{{Testa}, P.} et~al. (\bibinfo{year}{2020}).
\newblock \bibinfo{title}{{The Multi-slit Approach to Coronal Spectroscopy with
  the Multi-slit Solar Explorer (MUSE)}}.
\newblock {\it \bibinfo{journal}{\apj}\/},  {\it
  \bibinfo{volume}{888}\/}\bibinfo{issue}{(1)}, \bibinfo{pages}{3}.
  \DOIprefix\doi{10.3847/1538-4357/ab5b03}.
  \href{http://arxiv.org/abs/1909.08818}{\tt arXiv:1909.08818}.

\bibitem[{{De Pontieu} et~al.(2014){De Pontieu}, {Title}, {Lemen}, {Kushner},
  {Akin}, {Allard}, {Berger}, {Boerner}, {Cheung}, {Chou}, {Drake}, {Duncan},
  {Freeland}, {Heyman}, {Hoffman}, {Hurlburt}, {Lindgren}, {Mathur}, {Rehse},
  {Sabolish}, {Seguin}, {Schrijver}, {Tarbell}, {W{\"u}lser}, {Wolfson},
  {Yanari}, {Mudge}, {Nguyen-Phuc}, {Timmons}, {van Bezooijen}, {Weingrod},
  {Brookner}, {Butcher}, {Dougherty}, {Eder}, {Knagenhjelm}, {Larsen},
  {Mansir}, {Phan}, {Boyle}, {Cheimets}, {DeLuca}, {Golub}, {Gates}, {Hertz},
  {McKillop}, {Park}, {Perry}, {Podgorski}, {Reeves}, {Saar}, {Testa}, {Tian},
  {Weber}, {Dunn}, {Eccles}, {Jaeggli}, {Kankelborg}, {Mashburn}, {Pust},
  {Springer}, {Carvalho}, {Kleint}, {Marmie}, {Mazmanian}, {Pereira}, {Sawyer},
  {Strong}, {Worden}, {Carlsson}, {Hansteen}, {Leenaarts}, {Wiesmann},
  {Aloise}, {Chu}, {Bush}, {Scherrer}, {Brekke}, {Martinez-Sykora}, {Lites},
  {McIntosh}, {Uitenbroek}, {Okamoto}, {Gummin}, {Auker}, {Jerram}, {Pool} \&
  {Waltham}}]{2014SoPh..289.2733D}
\bibinfo{author}{{De Pontieu}, B.}, \bibinfo{author}{{Title}, A.~M.},
  \bibinfo{author}{{Lemen}, J.~R.} et~al. (\bibinfo{year}{2014}).
\newblock \bibinfo{title}{{The Interface Region Imaging Spectrograph (IRIS)}}.
\newblock {\it \bibinfo{journal}{\solphys}\/},  {\it
  \bibinfo{volume}{289}\/}\bibinfo{issue}{(7)}, \bibinfo{pages}{2733--2779}.
  \DOIprefix\doi{10.1007/s11207-014-0485-y}.
  \href{http://arxiv.org/abs/1401.2491}{\tt arXiv:1401.2491}.

\bibitem[{{Del Zanna} et~al.(2021){Del Zanna}, {Dere}, {Young} \&
  {Landi}}]{2021ApJ...909...38D}
\bibinfo{author}{{Del Zanna}, G.}, \bibinfo{author}{{Dere}, K.~P.},
  \bibinfo{author}{{Young}, P.~R.} et~al. (\bibinfo{year}{2021}).
\newblock \bibinfo{title}{{CHIANTI{\textemdash}An Atomic Database for Emission
  Lines. XVI. Version 10, Further Extensions}}.
\newblock {\it \bibinfo{journal}{\apj}\/},  {\it
  \bibinfo{volume}{909}\/}\bibinfo{issue}{(1)}, \bibinfo{pages}{38}.
  \DOIprefix\doi{10.3847/1538-4357/abd8ce}.
  \href{http://arxiv.org/abs/2011.05211}{\tt arXiv:2011.05211}.

\bibitem[{{Foley} et~al.(2001){Foley}, {Harra}, {Culhane} \&
  {Mason}}]{2001ApJ...560L..91F}
\bibinfo{author}{{Foley}, C.~R.}, \bibinfo{author}{{Harra}, L.~K.},
  \bibinfo{author}{{Culhane}, J.~L.} et~al. (\bibinfo{year}{2001}).
\newblock \bibinfo{title}{{Eruption of a Flux Rope on the Disk of the Sun:
  Evidence for the Coronal Mass Ejection Trigger?}}
\newblock {\it \bibinfo{journal}{\apjl}\/},  {\it
  \bibinfo{volume}{560}\/}\bibinfo{issue}{(1)}, \bibinfo{pages}{L91--L94}.
  \DOIprefix\doi{10.1086/324059}.

\bibitem[{{Gibson} \& {Fan}(2006)}]{2006JGRA..11112103G}
\bibinfo{author}{{Gibson}, S.~E.},  \& \bibinfo{author}{{Fan}, Y.}
  (\bibinfo{year}{2006}).
\newblock \bibinfo{title}{{Coronal prominence structure and dynamics: A
  magnetic flux rope interpretation}}.
\newblock {\it \bibinfo{journal}{Journal of Geophysical Research (Space
  Physics)}\/},  {\it \bibinfo{volume}{111}\/}\bibinfo{issue}{(A12)},
  \bibinfo{pages}{A12103}. \DOIprefix\doi{10.1029/2006JA011871}.

\bibitem[{{G{\"o}m{\"o}ry} et~al.(2016){G{\"o}m{\"o}ry}, {Veronig}, {Su},
  {Temmer} \& {Thalmann}}]{2016A&A...588A...6G}
\bibinfo{author}{{G{\"o}m{\"o}ry}, P.}, \bibinfo{author}{{Veronig}, A.~M.},
  \bibinfo{author}{{Su}, Y.} et~al. (\bibinfo{year}{2016}).
\newblock \bibinfo{title}{{Chromospheric evaporation flows and density changes
  deduced from Hinode/EIS during an M1.6 flare}}.
\newblock {\it \bibinfo{journal}{\aap}\/},  {\it \bibinfo{volume}{588}\/},
  \bibinfo{pages}{A6}. \DOIprefix\doi{10.1051/0004-6361/201527403}.
  \href{http://arxiv.org/abs/1602.02145}{\tt arXiv:1602.02145}.

\bibitem[{{Harra} \& {Sterling}(2003)}]{2003ApJ...587..429H}
\bibinfo{author}{{Harra}, L.~K.},  \& \bibinfo{author}{{Sterling}, A.~C.}
  (\bibinfo{year}{2003}).
\newblock \bibinfo{title}{{Imaging and Spectroscopic Investigations of a Solar
  Coronal Wave: Properties of the Wave Front and Associated Erupting
  Material}}.
\newblock {\it \bibinfo{journal}{\apj}\/},  {\it
  \bibinfo{volume}{587}\/}\bibinfo{issue}{(1)}, \bibinfo{pages}{429--438}.
  \DOIprefix\doi{10.1086/368079}.

\bibitem[{{Harra} et~al.(2011){Harra}, {Sterling}, {G{\"o}m{\"o}ry} \&
  {Veronig}}]{2011ApJ...737L...4H}
\bibinfo{author}{{Harra}, L.~K.}, \bibinfo{author}{{Sterling}, A.~C.},
  \bibinfo{author}{{G{\"o}m{\"o}ry}, P.} et~al. (\bibinfo{year}{2011}).
\newblock \bibinfo{title}{{Spectroscopic Observations of a Coronal Moreton
  Wave}}.
\newblock {\it \bibinfo{journal}{\apjl}\/},  {\it
  \bibinfo{volume}{737}\/}\bibinfo{issue}{(1)}, \bibinfo{pages}{L4}.
  \DOIprefix\doi{10.1088/2041-8205/737/1/L4}.

\bibitem[{{Imada} et~al.(2013){Imada}, {Aoki}, {Hara}, {Watanabe}, {Harra} \&
  {Shimizu}}]{2013ApJ...776L..11I}
\bibinfo{author}{{Imada}, S.}, \bibinfo{author}{{Aoki}, K.},
  \bibinfo{author}{{Hara}, H.} et~al. (\bibinfo{year}{2013}).
\newblock \bibinfo{title}{{Evidence for Hot Fast Flow above a Solar Flare
  Arcade}}.
\newblock {\it \bibinfo{journal}{\apjl}\/},  {\it
  \bibinfo{volume}{776}\/}\bibinfo{issue}{(1)}, \bibinfo{pages}{L11}.
  \DOIprefix\doi{10.1088/2041-8205/776/1/L11}.
  \href{http://arxiv.org/abs/1309.3401}{\tt arXiv:1309.3401}.

\bibitem[{{Innes} et~al.(2001){Innes}, {Curdt}, {Schwenn}, {Solanki},
  {Stenborg} \& {McKenzie}}]{2001ApJ...549L.249I}
\bibinfo{author}{{Innes}, D.~E.}, \bibinfo{author}{{Curdt}, W.},
  \bibinfo{author}{{Schwenn}, R.} et~al. (\bibinfo{year}{2001}).
\newblock \bibinfo{title}{{Large Doppler Shifts in X-Ray Plasma: An Explosive
  Start to Coronal Mass Ejection}}.
\newblock {\it \bibinfo{journal}{\apjl}\/},  {\it
  \bibinfo{volume}{549}\/}\bibinfo{issue}{(2)}, \bibinfo{pages}{L249--L252}.
  \DOIprefix\doi{10.1086/319164}.

\bibitem[{{Innes} et~al.(2003){Innes}, {McKenzie} \&
  {Wang}}]{2003SoPh..217..267I}
\bibinfo{author}{{Innes}, D.~E.}, \bibinfo{author}{{McKenzie}, D.~E.},  \&
  \bibinfo{author}{{Wang}, T.} (\bibinfo{year}{2003}).
\newblock \bibinfo{title}{{Observations of 1000 km s$^{-1}$ Doppler shifts in
  {}10$^{7}$ K solar flare supra-arcade}}.
\newblock {\it \bibinfo{journal}{\solphys}\/},  {\it
  \bibinfo{volume}{217}\/}\bibinfo{issue}{(2)}, \bibinfo{pages}{267--279}.
  \DOIprefix\doi{10.1023/B:SOLA.0000006874.31799.bc}.

\bibitem[{{Kleint} et~al.(2015){Kleint}, {Battaglia}, {Reardon}, {Sainz Dalda},
  {Young} \& {Krucker}}]{2015ApJ...806....9K}
\bibinfo{author}{{Kleint}, L.}, \bibinfo{author}{{Battaglia}, M.},
  \bibinfo{author}{{Reardon}, K.} et~al. (\bibinfo{year}{2015}).
\newblock \bibinfo{title}{{The Fast Filament Eruption Leading to the X-flare on
  2014 March 29}}.
\newblock {\it \bibinfo{journal}{\apj}\/},  {\it
  \bibinfo{volume}{806}\/}\bibinfo{issue}{(1)}, \bibinfo{pages}{9}.
  \DOIprefix\doi{10.1088/0004-637X/806/1/9}.
  \href{http://arxiv.org/abs/1504.00515}{\tt arXiv:1504.00515}.

\bibitem[{{Kohl} et~al.(1995){Kohl}, {Esser}, {Gardner}, {Habbal}, {Daigneau},
  {Dennis}, {Nystrom}, {Panasyuk}, {Raymond}, {Smith}, {Strachan}, {van
  Ballegooijen}, {Noci}, {Fineschi}, {Romoli}, {Ciaravella}, {Modigliani},
  {Huber}, {Antonucci}, {Benna}, {Giordano}, {Tondello}, {Nicolosi}, {Naletto},
  {Pernechele}, {Spadaro}, {Poletto}, {Livi}, {von der L{\"u}he}, {Geiss},
  {Timothy}, {Gloeckler}, {Allegra}, {Basile}, {Brusa}, {Wood}, {Siegmund},
  {Fowler}, {Fisher} \& {Jhabvala}}]{1995SoPh..162..313K}
\bibinfo{author}{{Kohl}, J.~L.}, \bibinfo{author}{{Esser}, R.},
  \bibinfo{author}{{Gardner}, L.~D.} et~al. (\bibinfo{year}{1995}).
\newblock \bibinfo{title}{{The Ultraviolet Coronagraph Spectrometer for the
  Solar and Heliospheric Observatory}}.
\newblock {\it \bibinfo{journal}{\solphys}\/},  {\it
  \bibinfo{volume}{162}\/}\bibinfo{issue}{(1-2)}, \bibinfo{pages}{313--356}.
  \DOIprefix\doi{10.1007/BF00733433}.

\bibitem[{{Lemen} et~al.(2012){Lemen}, {Title}, {Akin}, {Boerner}, {Chou},
  {Drake}, {Duncan}, {Edwards}, {Friedlaender}, {Heyman}, {Hurlburt}, {Katz},
  {Kushner}, {Levay}, {Lindgren}, {Mathur}, {McFeaters}, {Mitchell}, {Rehse},
  {Schrijver}, {Springer}, {Stern}, {Tarbell}, {Wuelser}, {Wolfson}, {Yanari},
  {Bookbinder}, {Cheimets}, {Caldwell}, {Deluca}, {Gates}, {Golub}, {Park},
  {Podgorski}, {Bush}, {Scherrer}, {Gummin}, {Smith}, {Auker}, {Jerram},
  {Pool}, {Soufli}, {Windt}, {Beardsley}, {Clapp}, {Lang} \&
  {Waltham}}]{2012SoPh..275...17L}
\bibinfo{author}{{Lemen}, J.~R.}, \bibinfo{author}{{Title}, A.~M.},
  \bibinfo{author}{{Akin}, D.~J.} et~al. (\bibinfo{year}{2012}).
\newblock \bibinfo{title}{{The Atmospheric Imaging Assembly (AIA) on the Solar
  Dynamics Observatory (SDO)}}.
\newblock {\it \bibinfo{journal}{\solphys}\/},  {\it
  \bibinfo{volume}{275}\/}\bibinfo{issue}{(1-2)}, \bibinfo{pages}{17--40}.
  \DOIprefix\doi{10.1007/s11207-011-9776-8}.

\bibitem[{{Long} et~al.(2013){Long}, {Williams}, {R{\'e}gnier} \&
  {Harra}}]{2013SoPh..288..567L}
\bibinfo{author}{{Long}, D.~M.}, \bibinfo{author}{{Williams}, D.~R.},
  \bibinfo{author}{{R{\'e}gnier}, S.} et~al. (\bibinfo{year}{2013}).
\newblock \bibinfo{title}{{Measuring the Magnetic-Field Strength of the Quiet
  Solar Corona Using ``EIT Waves''}}.
\newblock {\it \bibinfo{journal}{\solphys}\/},  {\it
  \bibinfo{volume}{288}\/}\bibinfo{issue}{(2)}, \bibinfo{pages}{567--583}.
  \DOIprefix\doi{10.1007/s11207-013-0331-7}.
  \href{http://arxiv.org/abs/1305.5169}{\tt arXiv:1305.5169}.

\bibitem[{{Pike} \& {Mason}(2002)}]{2002SoPh..206..359P}
\bibinfo{author}{{Pike}, C.~D.},  \& \bibinfo{author}{{Mason}, H.~E.}
  (\bibinfo{year}{2002}).
\newblock \bibinfo{title}{{EUV Spectroscopic Observations of Spray Ejecta from
  an X2 Flare}}.
\newblock {\it \bibinfo{journal}{\solphys}\/},  {\it
  \bibinfo{volume}{206}\/}\bibinfo{issue}{(2)}, \bibinfo{pages}{359--381}.
  \DOIprefix\doi{10.1023/A:1015093902578}.

\bibitem[{{Schrijver} et~al.(2011){Schrijver}, {Aulanier}, {Title}, {Pariat} \&
  {Delann{\'e}e}}]{2011ApJ...738..167S}
\bibinfo{author}{{Schrijver}, C.~J.}, \bibinfo{author}{{Aulanier}, G.},
  \bibinfo{author}{{Title}, A.~M.} et~al. (\bibinfo{year}{2011}).
\newblock \bibinfo{title}{{The 2011 February 15 X2 Flare, Ribbons, Coronal
  Front, and Mass Ejection: Interpreting the Three-dimensional Views from the
  Solar Dynamics Observatory and STEREO Guided by Magnetohydrodynamic Flux-rope
  Modeling}}.
\newblock {\it \bibinfo{journal}{\apj}\/},  {\it
  \bibinfo{volume}{738}\/}\bibinfo{issue}{(2)}, \bibinfo{pages}{167}.
  \DOIprefix\doi{10.1088/0004-637X/738/2/167}.

\bibitem[{{Tian} et~al.(2012){Tian}, {McIntosh}, {Xia}, {He} \&
  {Wang}}]{2012ApJ...748..106T}
\bibinfo{author}{{Tian}, H.}, \bibinfo{author}{{McIntosh}, S.~W.},
  \bibinfo{author}{{Xia}, L.} et~al. (\bibinfo{year}{2012}).
\newblock \bibinfo{title}{{What can We Learn about Solar Coronal Mass
  Ejections, Coronal Dimmings, and Extreme-ultraviolet Jets through
  Spectroscopic Observations?}}
\newblock {\it \bibinfo{journal}{\apj}\/},  {\it
  \bibinfo{volume}{748}\/}\bibinfo{issue}{(2)}, \bibinfo{pages}{106}.
  \DOIprefix\doi{10.1088/0004-637X/748/2/106}.
  \href{http://arxiv.org/abs/1201.2204}{\tt arXiv:1201.2204}.

\bibitem[{{Veronig} et~al.(2011){Veronig}, {G{\"o}m{\"o}ry}, {Kienreich},
  {Muhr}, {Vr{\v{s}}nak}, {Temmer} \& {Warren}}]{2011ApJ...743L..10V}
\bibinfo{author}{{Veronig}, A.~M.}, \bibinfo{author}{{G{\"o}m{\"o}ry}, P.},
  \bibinfo{author}{{Kienreich}, I.~W.} et~al. (\bibinfo{year}{2011}).
\newblock \bibinfo{title}{{Plasma Diagnostics of an EIT Wave Observed by
  Hinode/EIS and SDO/AIA}}.
\newblock {\it \bibinfo{journal}{\apjl}\/},  {\it
  \bibinfo{volume}{743}\/}\bibinfo{issue}{(1)}, \bibinfo{pages}{L10}.
  \DOIprefix\doi{10.1088/2041-8205/743/1/L10}.
  \href{http://arxiv.org/abs/1111.3505}{\tt arXiv:1111.3505}.

\bibitem[{{Wang} et~al.(2007){Wang}, {Sui} \& {Qiu}}]{2007ApJ...661L.207W}
\bibinfo{author}{{Wang}, T.}, \bibinfo{author}{{Sui}, L.},  \&
  \bibinfo{author}{{Qiu}, J.} (\bibinfo{year}{2007}).
\newblock \bibinfo{title}{{Direct Observation of High-Speed Plasma Outflows
  Produced by Magnetic Reconnection in Solar Impulsive Events}}.
\newblock {\it \bibinfo{journal}{\apjl}\/},  {\it
  \bibinfo{volume}{661}\/}\bibinfo{issue}{(2)}, \bibinfo{pages}{L207--L210}.
  \DOIprefix\doi{10.1086/519004}. \href{http://arxiv.org/abs/0709.2329}{\tt
  arXiv:0709.2329}.

\bibitem[{{Warwick}(1957)}]{1957ApJ...125..811W}
\bibinfo{author}{{Warwick}, J.~W.} (\bibinfo{year}{1957}).
\newblock \bibinfo{title}{{Flare-Connected Prominences.}}
\newblock {\it \bibinfo{journal}{\apj}\/},  {\it \bibinfo{volume}{125}\/},
  \bibinfo{pages}{811}. \DOIprefix\doi{10.1086/146354}.

\bibitem[{{Watanabe} et~al.(2010){Watanabe}, {Hara}, {Sterling} \&
  {Harra}}]{2010ApJ...719..213W}
\bibinfo{author}{{Watanabe}, T.}, \bibinfo{author}{{Hara}, H.},
  \bibinfo{author}{{Sterling}, A.~C.} et~al. (\bibinfo{year}{2010}).
\newblock \bibinfo{title}{{Production of High-temperature Plasmas During the
  Early Phases of a C9.7 Flare}}.
\newblock {\it \bibinfo{journal}{\apj}\/},  {\it
  \bibinfo{volume}{719}\/}\bibinfo{issue}{(1)}, \bibinfo{pages}{213--219}.
  \DOIprefix\doi{10.1088/0004-637X/719/1/213}.

\bibitem[{{Young}(2022)}]{2022zndo...6339584Y}
\bibinfo{author}{{Young}, P.~R.} (\bibinfo{year}{2022}).
\newblock {\it \bibinfo{title}{{EIS\_AUTO\_FIT and SPEC\_GAUSS\_EIS: Gaussian
  fitting routines for the Hinode/EIS mission}}\/}.
\newblock \bibinfo{type}{Technical Report}.
\newblock \DOIprefix\doi{10.5281/zenodo.6339584}.

\bibitem[{{Young} et~al.(2016){Young}, {Dere}, {Landi}, {Del Zanna} \&
  {Mason}}]{2016JPhB...49g4009Y}
\bibinfo{author}{{Young}, P.~R.}, \bibinfo{author}{{Dere}, K.~P.},
  \bibinfo{author}{{Landi}, E.} et~al. (\bibinfo{year}{2016}).
\newblock \bibinfo{title}{{The CHIANTI atomic database}}.
\newblock {\it \bibinfo{journal}{Journal of Physics B Atomic Molecular
  Physics}\/},  {\it \bibinfo{volume}{49}\/}\bibinfo{issue}{(7)},
  \bibinfo{pages}{074009}. \DOIprefix\doi{10.1088/0953-4075/49/7/074009}.
  \href{http://arxiv.org/abs/1512.05620}{\tt arXiv:1512.05620}.

\bibitem[{{Young} et~al.(2013){Young}, {Doschek}, {Warren} \&
  {Hara}}]{2013ApJ...766..127Y}
\bibinfo{author}{{Young}, P.~R.}, \bibinfo{author}{{Doschek}, G.~A.},
  \bibinfo{author}{{Warren}, H.~P.} et~al. (\bibinfo{year}{2013}).
\newblock \bibinfo{title}{{Properties of a Solar Flare Kernel Observed by
  Hinode and SDO}}.
\newblock {\it \bibinfo{journal}{\apj}\/},  {\it
  \bibinfo{volume}{766}\/}\bibinfo{issue}{(2)}, \bibinfo{pages}{127}.
  \DOIprefix\doi{10.1088/0004-637X/766/2/127}.
  \href{http://arxiv.org/abs/1212.4388}{\tt arXiv:1212.4388}.

\end{thebibliography}





\end{document}